\documentclass[prb,superscriptaddress,showpacs]{revtex4-1}

\usepackage{graphicx,bm,color}

\usepackage{graphicx}% Include figure files 
\usepackage{dcolumn}% Align table columns on decimal point  
\usepackage{bm}% bold math
\usepackage{mhchem}
\usepackage{epstopdf}
\usepackage{color}

\begin{document}

% Be sure to use the \title, \author, \affiliation, and \abstract macros
% to format your title page.  Don't use lower-level macros to  manually
% adjust the fonts and centering.

\title{Anisotropic Heisenberg form of RKKY interaction in the one-dimensional spin-polarized electron gas}
%\title{Beating Pattern in the RKKY Interaction for the Spin Polarized Electron Gas }

\author{Mohammad M. Valizadeh}
\email{mvbr5@mail.missouri.edu}

\affiliation{Department of Physics $\&$ Astronomy, University of Missouri, Columbia, MO 65211, USA}

%\author{S. Satpathy}
%\email{satpathys@missouri.edu}

%\affiliation{Department of Physics $\&$ Astronomy, University of Missouri, Columbia, MO 65211, USA}
% Please provide a full mailing address here. 

%\author{David P. Jackson} % optional
%\email{ajp@dickinson.edu} % optional
%\affiliation{Department of Physics, Dickinson College, Carlisle, PA 17013} % optional

% See the REVTeX documentation for more examples of author and affiliation lists.

\date{\today}

\begin{abstract}

We study the indirect exchange interaction between two localized magnetic moments, known as Ruderman-Kittel-Kasuya-Yosida (RKKY) interaction, in a one-dimensional spin-polarized electron gas. 
We find explicit expressions for each term of this interaction, study their oscillatory behaviors as a function of the distance between two magnetic moments, $R$, and compare them with the
known results for RKKY interaction in the case of one-dimensional standard electron gas. 
We show this interaction can be written in an anisotropic Heisenberg form, 
$E(\vec R)=\lambda^2\chi_{xx}(S_{1x}S_{2x}+S_{1y}S_{2y})+\lambda^2\chi_{zz}\, S_{1z}S_{2z}$, coming from broken time-reversal symmetry of the host material.

\end{abstract}
% AJP requires an abstract for all regular article submissions.
% Abstracts are optional for submissions to the "Notes and Discussions" section.

\maketitle % title page is now complete

\section{Introduction}

The well-known RKKY interaction between two localized magnetic moments in a host solid 
was originally derived for the free electron gas
% in the seminal works of Ruderman, Kittel, Kasuya, and Yosida
\cite{RKKY1,RKKY2,RKKY3} and continues to be an active and important research field in different classes of materials with different symmetries \cite{Liechtenstein,Bruno,Asgari2,Asgari,Chesi,Fransson,Sherafati}. 
Straightforward calculations lead to the oscillatory Heisenberg form of RKKY interaction, $ J  \vec S_1 \cdot \vec S_2$, in a two- and three-dimensional electron gas, 
but the result which provides the correct answer to the same problem in a one-dimensional (1D) electron gas
was always challenging \cite{1DKittel,1D1,1D2} due to the dependence of the double intergral on the order of integration.
Similar difficulties appear for the case of the one-dimensional spin polarized electron gas.
Effect of the broken time-reversal symmetry on the RKKY interaction has been studied in two- and three-dimensions \cite{spin-polarized1,spin-polarized2}.
In this brief manuscript, we study this effect for the case of one-dimensional electron gas, which provides a useful addition to the literature.

In the presence of spin-orbit coupling, Dzyaloshinsky and Moriya (DM) have used the lattice models to show \cite{DM,Moriya}
%there appears a vector or a tensor interaction term, $\vec D$ 
%and $ \stackrel{\leftrightarrow}{\Gamma}$, respectively, 
vector and tensor interaction terms appear in addition to the scalar RKKY interaction, so the $\vec R$-dependent part of the change in the energy of the system after adding two localized magnetic moments
can be written as
$ E(\vec R) = J \vec S_1 \cdot \vec S_2 + \vec D \cdot   \vec S_1 \times \vec S_2 
+  \vec S_1 \cdot  \stackrel{\leftrightarrow}{\Gamma} \cdot \vec S_2    $.
For the case of spin-polarized electron gas, due to the presence of inversion symmetry, one may expect $\vec D=\vec 0$, but broken time reversal symmetry leads to an extra term which has the form
of tensor DM interaction, $\vec S_1 \cdot  \stackrel{\leftrightarrow}{\Gamma} \cdot \vec S_2$. This term causes the anisotropy Heisenberg-form of the interaction and beat-pattern in their oscillatory behaviors.
Studying of the RKKY and DM interactions in a host solid with broken symmetries plays a crucial role to understand the physics of the magnetic chiral ordering and Skyrmions \cite{Tokura,sk1,sk2,skyrmea1962}.
\section{General formalism for $\vec R$-dependent part of energy}
The second localized magnetic moment, $\vec S_2$, located at $\vec r=\vec R$, interacts with the wave functions of the electrons perturbed by the first localized magnetic moment, $\vec S_1$, which is located at the origin, $\vec r=0$.
The net interaction can be written as $E = E_{0} + E(\vec R)$, where $E_0$ is a constant shift in the energy of the system and can be evaluated using the first order perturbation theory.
Defining the contact interaction between the localized moments
%, $\vec S_1$ and $\vec S_2$ located at $\vec r=\vec 0$ and $\vec r=\vec R$ respectively, 
and the spin of electrons in the operator forms of  $\hat V_1 =-\lambda     \sum_{\sigma \sigma ^\prime}   |\vec 0 \sigma \rangle  \vec S_1 \cdot \vec s \     \langle \vec 0 \sigma^\prime |$
and $\hat V_2 =-\lambda     \sum_{\sigma \sigma ^\prime}   |\vec R \sigma \rangle  \vec S_2 \cdot \vec s \     \langle \vec R \sigma^\prime |$,
and using Lippmann-Schwinger equation\cite{Lippmann-Schwinger}, 
one can find the $\vec R$-dependent part of this interaction as
\begin{equation}
E(\vec R)= \frac{-\lambda^2}{\pi}\,\,Im \int_{-\infty}^{E_F} Tr\big[G(0,\vec R, E)\, \vec S_2\cdot\vec s\, G(\vec R, 0, E) \vec S_1\cdot\vec s\, \big]dE.
\label{master}
\end{equation}
Here, $\lambda$ is a phenomenological constant responsible for
the strength of the contact interaction between localized magnetic moments and the spin of electrons,
and $G(\vec R,0, E)$ and $G(0,\vec R, E)$ are the unperturbed retarded Green's function matrices which their elements can be found using
\begin{equation}
G_{\sigma_1\sigma_2}(\vec r_1,\vec r_2,E)=\sum\limits_{\vec k\nu}^{\infty}\frac{\psi_{\vec k\nu}(\vec r_1,\sigma_1)\psi_{\vec k\nu}^*(\vec r_2,\sigma_2)}{E+i\mu-\varepsilon_{\vec k\nu}},
\label{GF}
\end{equation}
with $\mu\rightarrow 0^+$, and $\nu$ as the band index. 
Although Eq. (\ref{master}) is a known formula and has been widely used in the literature \cite{Bruno,Asgari2,Asgari,spin-polarized2},
working through its derivation still holds value.
A pedagogical derivation of this equation, which is given in the Appendix, provides some important information about limitation of using this equation to get the indirect exchange interaction
between two localized magnetic moments, and 
can be helpful for both experts and beginners in this research field. 
A very essential point that can be extracted is to get the result as Eq. (\ref{master}), the order of integration between $E$ and $\vec k$, immediately after Eq. (\ref{ap5}) in the Appendix, has been changed.
If the Fubini's condition \cite{Fubini1,Fubini2} is not satisfied, changing the order of integration in a double-intergral can lead to a different result. 
For the case of one-dimensional electron gas, one can immediately find that Fubini's condition is not satisfied.

\section{Green's function matrices} 

For the case of one-dimensional spin-polarized electron gas the unperturbed eigenstates and energy bands are defined as $| \vec k \sigma    \rangle = \frac{1}{\sqrt L} e^{i \vec k \cdot \vec r} |\sigma \rangle$
and $\varepsilon_{\vec k \sigma}    =  \frac{\hbar^2 k^2}{2m} \mp \Delta$, respectively, 
where we have assumed $+(-)$ sign for spin-down (-up) states. 
To find the elements of Green's function matrix, one can use Eq. (\ref{GF}) and find the result

\begin{equation}
 G(\vec r_1, \vec r_2,E)
 =-\frac{im}{\hbar^{2}}
\begin{pmatrix}
\frac {e^{i\alpha(E+\Delta)r}}{\alpha(E+\Delta)} & 0 \\
0 & \frac {e^{i\alpha(E-\Delta)r}}{\alpha(E-\Delta)}  
 \end{pmatrix},
 \label{masoud}
\end{equation}
where
\begin{eqnarray}
\alpha (x) =\begin{cases}
 (2m\hbar^{-2} x)^{1/2}  \text { if  $ x > 0$},   \\
i  (2m\hbar^{-2}|x|)^{1/2}  \text { if  $ x < 0$}.
 \label{eq5}
\end{cases}
\end{eqnarray}
For this case, $G_{\sigma\sigma'}(\vec R,0,E)=G_{\sigma\sigma'}(0,\vec R,E)$. The last result could be expected due to the presence of inversion symmetry.
If we set $\Delta=0$, we can easily find the results for the Green's function matrices for the case of standard one-dimensional electron gas. 

{\it Discussions} --
Putting the Green's function matrices found in Eq. (3) into the Eq. (\ref{master}) and applying standard complex-plane integration techniques lead to the wrong results for the RKKY interaction.
For example, for the case of standard one-dimensional electron gas, it leads to the wrong unphysical answer to the problem, $E(\vec R)=J\,\vec S_1\cdot \vec S_2$, with
\begin{equation}
J=\frac{\lambda^2\,m}{2\pi}{\rm Si}(2 k_{F}R),
\label{eq8}	
\end{equation}
where ${\rm Si}(x)$ is the sine intergral,
\begin{equation}
{\rm Si}(x)=\int_{0}^{x}\frac{{\rm sin}(t)}{t}dt.    
\label{wrong}	
\end{equation}
Here, we would like to emphasize using Eq. (\ref{master}) with the Green's function matrices found in Eq. (3) leads to the wrong unphysical results for both the cases of
standard and spin-polarized one-dimensional electron gas.
In a similar work \cite{Bruno}, Imamura $et\,\, al.$ have used the same method. To obtain physical answer to the problem, 
the authors inserted a constant, $-\lambda^2m/2\pi$ found by Yafet \cite{1D1}, to the last result. 
Yafet explained \cite{1D1} that the strong singularities in the intergrant of double integral is responsible for the dependence of the double integral 
on the order of integration. Adding this constant to last result for $J$ leads to the right answer to the problem.

On the other hand, Giuliani and coauthors\cite{1D2} provided a very pedagogical method to obtain the right answer to the problem. 
This method is based on finding the delocalized eigenstates of the one-dimensional Schrodinger equation with a delta-fucntion as the potential, $V(x)=u\,\delta(x)$ when $u\rightarrow 0^+$, and then using them to find the electronic density.
Similarly, one can use these eigenstates
to find the full Green's function of the system as $G_F( r,r\prime,E)=g( r,r\prime,E)\,\sigma_0$, where
\begin{align}
g( r,r\prime,E)=&\frac{1}{2\pi}\bigg\{  \int_{0}^{\infty} \frac{\rm{cos} (k\mid r-r'\mid)}{E+i\mu-\frac{\hbar^2k^2}{2m}}dk  \nonumber \\
                                     - &  \int_{0}^{\infty}\frac{2u^2}{4k^2+u^2} \frac{\rm{cos} (k\mid r+r'\mid)}{E+i\mu-\frac{\hbar^2k^2}{2m}}dk  \nonumber \\
                                     + &  \int_{0}^{\infty}\frac{4ku}{4k^2+u^2} \frac{\rm{sin} (k\mid r+r'\mid)}{E+i\mu-\frac{\hbar^2k^2}{2m}}dk \bigg\}.    
\label{fullG}
\end{align}
The last result is valid for any value of $u$. Now using the full and unperturbed Green's function matrices, it is convenient to find the change in the local electronic density, $\delta n(x)=\frac{-1}{\pi}Im\int_{0}^{E_F}Tr[G_F(x,x,E)-G(x,x,E)]dE$, 
for the case that $u\rightarrow0^+$.
The result provides the right answer to the problem, viz.
\begin{equation}
\delta n(x)=-\frac{u}{\pi}\,{\rm si}(2k_F|x|),
\label{1dres}	
\end{equation}
where ${\rm si}(x)={\rm Si}(x)-\frac{\pi}{2}$. 
To get the last result, one may need to use the following Dirac delta function equalities 
\begin{align}
           &\delta(k)=\frac{2}{\pi}\lim_{u \to 0^+}\frac{u}{4k^2+u^2},  \nonumber \\
           & \delta(E)=-\frac{1}{\pi}\lim_{\mu \to 0^+}Im[\frac{1}{E+i\mu}], \nonumber \\
            \delta(k^2-\frac{2mE}{\hbar^2})=&\sqrt{\frac{\hbar^2}{8mE}}\bigg\{ \delta(k-\sqrt{\frac{2mE}{\hbar^2}})+\delta(k+\sqrt{\frac{2mE}{\hbar^2}}) \bigg\}.    
\label{fullG}
\end{align}
Similar methods can be used to find the magnetic interaction between two localized moments in a spin-polarized one-dimensional electron gas.

\section{One-dimensional Spin-polarized electron gas} 

Writing the interaction in the form of $E=E_0+E(\vec R)$, where $E_0$ is a constant shift to the energy of the system and $E(\vec R)$ is the distance-dependent part of the interaction,
and using the first-order perturbation theory, one can easily find
\begin{align}
  E_0=&\sum\limits_{\vec k\nu}^{occ} \langle\vec k\nu\mid\hat V_1+\hat V_2\mid\vec k\nu\rangle,     \nonumber \\
     =&-\frac{\lambda\hbar}{2\pi}(S_{1z}+S_{2z})(k_{F\uparrow}-k_{F\downarrow}).
\label{E0}	
\end{align}
Choosing the $z$-direction as the quantization axis leads to this non-zero $E_0$ for spin-polarized electron gas.
For the case of standard one-dimensional electron gas, $k_{F\uparrow}=k_{F\downarrow}$ and $E_0=0$.

Using similar methods, and defining $\delta=\sqrt{2m\hbar^{-2}\Delta}$, we 
found the $\vec R$-dependent part of the interaction between two localized magnetic moments in the one-dimensional spin-polarized electron gas has the form of
$E(\vec R)=J\vec S_1\cdot   \vec S_2+\vec S_1 \cdot  \stackrel{\leftrightarrow}{\Gamma}  \cdot  \vec S_2$, where
\begin{equation}
J=\frac{\lambda^2\,m}{4\pi}\{{\rm si}(2 k_{F\uparrow}R)+{\rm si}(2k_{F\downarrow}R)\},
\label{eq8}	
\end{equation}
and the survived tensor interaction has the form of:
\begin{equation}
\stackrel{\leftrightarrow}{\Gamma} =
\begin{pmatrix}
P & 0 & 0\\
0 & P & 0\\
0 & 0 & 0
 \end{pmatrix},
 \label{eq10}
\end{equation}
where 
\begin{align}
  P=&\frac{\lambda^2\,m}{2\pi}\,\bigg(
                  - \int_{0}^{\sqrt{2}\delta} \frac{e^{-\sqrt{2\delta^2-k^2}R} {\rm cos}[kR]}{\sqrt{2\delta^2-k^2}} dk     \nonumber \\
                  + &\int_{\sqrt{2}\delta}^{k_{F\uparrow}} \frac{{\rm sin}[\sqrt{k^2-2\delta^2}R] {\rm cos}[kR]}{\sqrt{k^2-2\delta^2}} dk    \nonumber \\
                  + & \int_{0}^{k_{F\downarrow}}\frac{{\rm sin}[\sqrt{k^2+2\delta^2}R] {\rm cos}[kR]}{\sqrt{k^2+2\delta^2}} dk -g \bigg),
\label{eq11}	
\end{align}
with
\begin{eqnarray}
g =\begin{cases}
 J \,\,\,\,\,\,\,\,\,\,\,\,\,\,\,\,\,\,\,         \text { if  $\delta\neq0$},   \\
 J+\frac{\pi}{2} \,\,\,\,\,\,\,\, \text { if  $\delta =0$}.
 \label{eq12}
\end{cases}
\end{eqnarray}
%
%
%: Fig. 3
\begin{figure}
\includegraphics[angle=0,width=0.49    \linewidth]{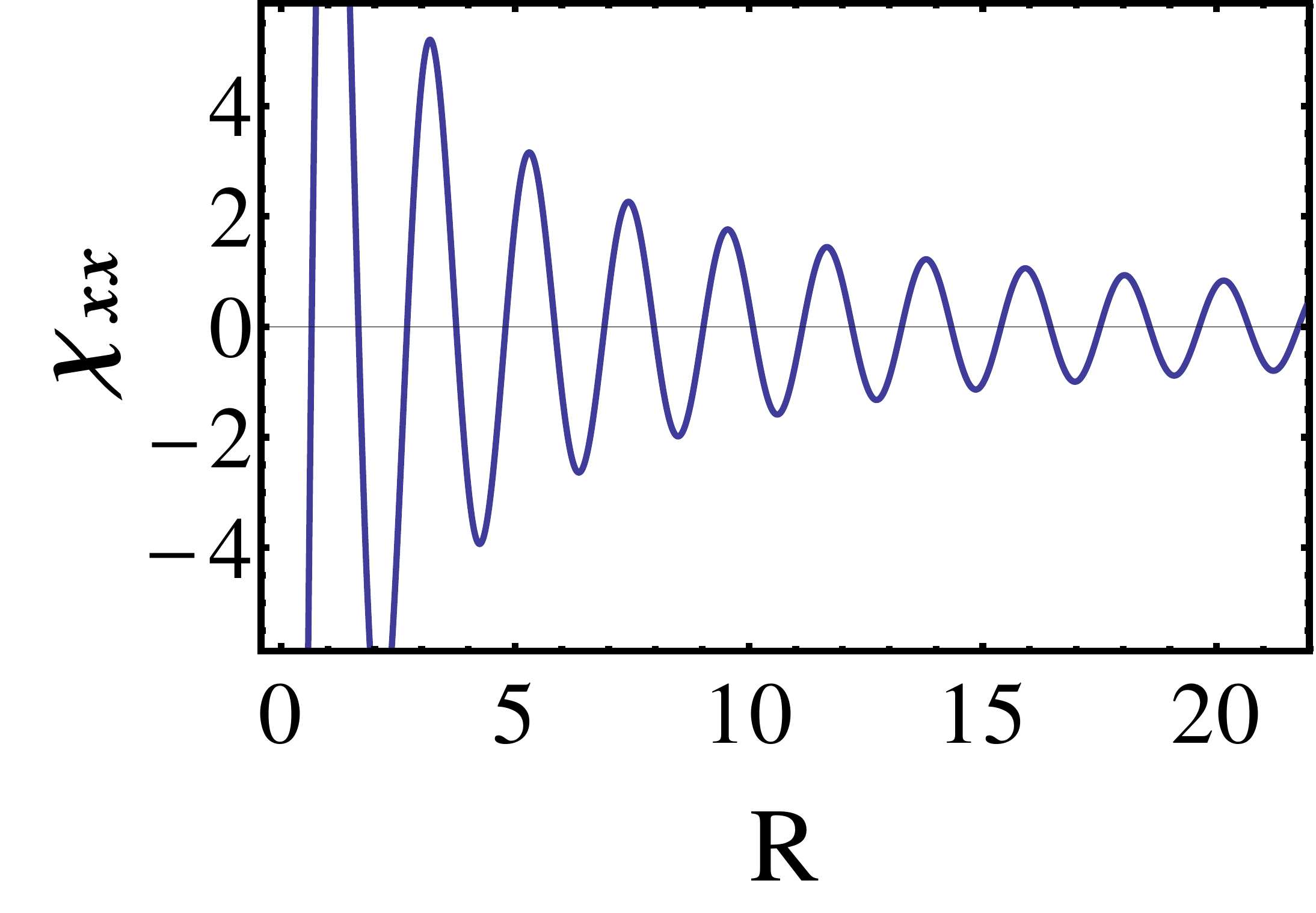}
\hfill
\includegraphics[angle=0,width=0.50    \linewidth]{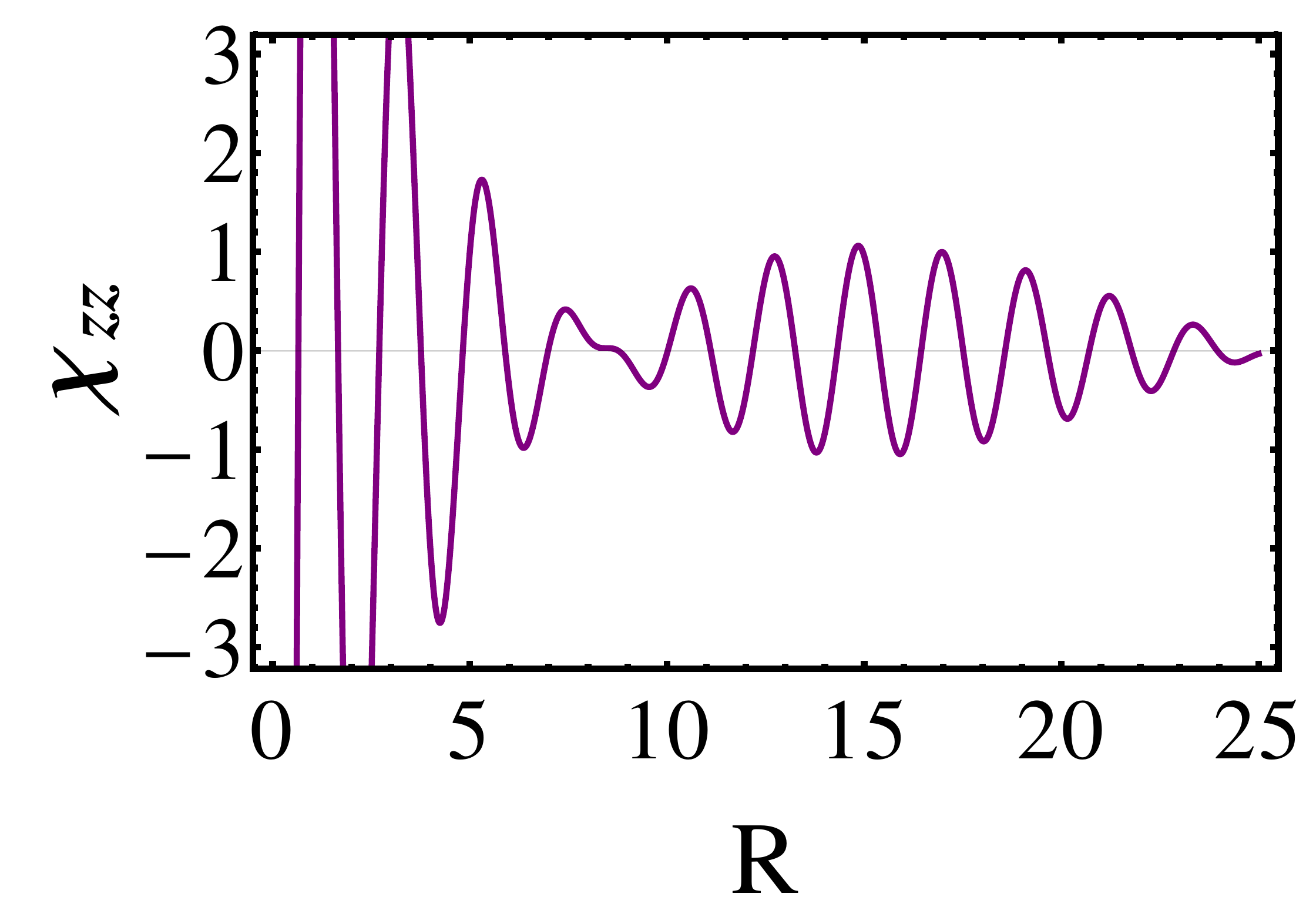}
\caption{Oscillatory behaviors of   
$\chi_{zz}$ ({\it right}) and $\chi_{xx}$ ({\it left}) in units of    
$10^{-2} m\pi^{-1}$ as a function of $R$ in the units of \AA\ for given values of $k_{F\uparrow}=1.574$ and $k_{F\downarrow}=1.388$ 1/\AA\.
The beat-pattern oscillatory behavior of $\chi_{zz}$ can be easily understood based on two different Fermi momenta, $k_{F\uparrow}$ and $k_{F\downarrow}$, in the system.
}
\label{fig1}
\end{figure}

One can immediately learn the vector DM, $\vec D$, vanishes. This result could be physically expected, based on the lack of explicit spin-orbit coupling and the presence of inversion symmetry in the system. 
Note that for an unpolarized electron gas $\Delta=0$, so $\delta=0$ which leads to $k_{F\uparrow} = k_{F\downarrow}$, and finally, above expressions lead to $P=0$ and the well-known $J$ for the case of one-dimensional electron gas \cite{spin-polarized1,spin-polarized2,Bruno}.
Adding all of the energy terms, it is convenient to find an anisotropic Heisenberg form for this interaction, 
\begin{equation}
E(\vec R)=\lambda^2\chi_{xx}(S_{1x}S_{2x}+S_{1x}S_{2x})+\lambda^2\chi_{zz}\, S_{1z}S_{2z},
\label{eq13}	
\end{equation}
with $\chi_{xx}=(P+J^{1D})/\lambda^2$ and $\chi_{zz}=J^{1D}/\lambda^2$. Oscillatory behaviors of $\chi_{xx}$ and $\chi_{zz}$ as a function of $R$, have been studied in the Fig. (\ref{fig1}).

To study the long-range behavior of $\chi_{zz}$, one can write the sine integral in terms of auxiliary functions, ${\rm Si}(x)\simeq\frac{\pi}{2}-f(x){\rm cos}(x)-g(x){\rm sin}(x)$, then use asymptotic series,
\begin{align}
& f(x)\simeq\frac{1}{x}\big(1-\frac{2!}{x^2}+\frac{4!}{x^4}-\frac{6!}{x^6} + ... \big),    \nonumber \\
& g(x)\simeq\frac{1}{x^2}\big(1-\frac{3!}{x^2}+\frac{5!}{x^4}-\frac{7!}{x^6} + ... \big),
\label{eq14}
\end{align}
so we will get ${\rm si}(x)\simeq -\,{\rm cos}(x)/x$ which leads to the result
\begin{equation}
\chi_{zz}\simeq-\,\frac{m}{4\pi}\big(\frac{{\rm cos}(2k_{F\uparrow} R)}{2k_{F\uparrow} R}+\frac{{\rm cos}(2k_{F\downarrow} R)}{2k_{F\downarrow} R} \big),
\label{eq15}	
\end{equation}
for large distances, $R\gg \,\,k_{F\uparrow}^{-1}$ and $k_{F\downarrow}^{-1}$. This result is compatible with the well-known long-range behavior of RKKY interaction in $n$-dimension, viz.,
\begin{equation}
\chi(\vec R)\propto\frac{{\rm cos}\big(2k_{F}R-\frac{(n-1)\pi}{2}\big)}{(2k_{F}R)^n}.
\label{eq16}	
\end{equation}
The last result is different from what Sobota $et\,\, al.$ \cite{Sobota} reported and provides right answer for the long-distance behavior of RKKY interaction in $n$-dimension. 

It is worthwhile to mention that, disregarding the coefficients and the decay-factor $R^{-n}$, 
the long-distance behaviors of $J$ for the cases of one- and three-dimensional spin-polarized electron gas show a similar pattern, 
viz., $J\propto{\rm cos}(k_{F\uparrow}R+k_{F\downarrow}R)\times {\rm cos}(k_{F\uparrow}R-k_{F\downarrow}R)$, 
while for the case of two-dimensional spin-polarized electron gas, ${\rm sin}(k_{F\uparrow}R+k_{F\downarrow}R)\times {\rm cos}(k_{F\uparrow}R-k_{F\downarrow}R)$ 
form of oscillatory behavior is expected. 
This fact can be found by comparing Eq. (\ref{eq15}) with the final results for long-distance behaviors of $J$ in a two- and three-dimensional spin-polarized electron gas\cite{spin-polarized1,spin-polarized2}, and
is valid for the case that $\delta\ll \,\,k_{F\uparrow}$ and $k_{F\downarrow}$.

\section{Summary} 

We found an anisotropic Heisenberg form for the indirect exchange interaction between two localized magnetic moments in a one-dimensional spin-polarized electron gas.
Explicit expressions are found for each term, and long-range behaviors of them have been studied. 
In this case, ${\rm cos}(k_{F\uparrow}R-k_{F\downarrow}R)$ and ${\rm cos}(k_{F\uparrow}R+k_{F\downarrow}R)$ control the long-distance beat-pattern and oscillatory behavior of $J$, respectively,
and the magnetic interaction falls off as $R^{-1}$. 
Disregarding the coefficients and the decay-factor $R^{-n}$, the same argument is valid for the case of three-dimensional spin-polarized electron gas. 
Setting $k_{F\uparrow} = k_{F\downarrow}$ leads to the standard result for the RKKY interaction $J$ for the one-dimensional electron gas.
Numerical results showed a beating pattern due to the interference between the two Fermi momenta $k_{F\uparrow}$ and  $k_{F\downarrow}$ for the two different spin channels. 
\section*{Acknowledgements}

The author would like to 
thank Sashi Satpathy, Giovanni Vignale and Brett A. Heischmidt for helpful discussions.
This research is supported by the U.S. Department of Energy, Office of Basic Energy Sciences, Division of Materials Sciences and Engineering for financial support under Award    No.DE-FG02-00ER45818.  

\section{Appendix} 
Here, we derive a general expression for the interaction between two localized magnetic moments in a host material, known as RKKY and Dzyaloshinsky-Moriya interactions, 
using Lippmann-Schwinger equation\cite{Lippmann-Schwinger}. Although the final result of this Appendix, Eq. (\ref{master}), is known in the literature, working through this derivation is worthwhile.
It can be useful for both beginners and experts in the field, and at the same time it provides some important information about the final result and its limitations. 

First, we put the first localized magnetic moment $\vec S_1$  at $\vec r_1=\vec 0$ and it perturbs the state of electrons, then second localized magnetic moment 
$\vec S_2$ will be located at $\vec r_2=\vec R$ and interacts with the perturbed electrons of the host. 
%
%\begin{figure}[h!]
%\centering
%\includegraphics[width=4in]{RKKY-Masoud.pdf}
% Notice the width specification.  Photographs should normally have a
% resolution of approximately 300 pixels per inch when printed, that is,
% a total width of about 1000 pixels for a photo to be printed one column
% wide.  Note also that this included photo is in .jpg format even though 
% a .tiff version should be submitted for final production.
%\caption{We put the first and second localized magnetic moments at $\vec r_1=0$ and $\vec r_2=\vec R$, respectively.}
%label{fig1}
%\end{figure}
%
%
Showing the perturbed and unperturbed states by $\mid\widetilde{k\nu}\rangle$ and $\mid k\nu\rangle$, respectively, one can use Lippmann-Schwinger equation to find
\begin{equation}
\mid \widetilde{k\nu}\rangle\simeq\mid k\nu\rangle+\hat G\hat V_1\mid k\nu\rangle,
\label{ap1}
\end{equation}
where $\nu$ is the band index, $\hat G(E)=(E+i\mu-\hat H)^{-1}$ with $\mu\rightarrow 0^+$, is the unperturbed retarded Green's function operator to be evaluated at $E=\varepsilon_{k\nu}$, 
and $\hat V_i$ indicates the perturbing potential operator applied by the $i$-th localized magnetic moment, and can be written as
\begin{equation}
\hat V_i=-\lambda\sum\limits_{\sigma\sigma'}\mid\,\vec r_i\,\sigma\rangle\langle\vec r_i\,\sigma'\mid\vec S_i\cdot\vec s_{\sigma\sigma'}.
\label{ap2}
\end{equation}
This perturbing potential has the contact interaction forms of $V_1(\vec r\,)=-\lambda\, \delta(\vec r\,)\,\vec S_1\cdot\vec s$ and $V_2(\vec r\,)=-\lambda\, \delta(\vec r-\vec R)\,\vec S_2\cdot\vec s$. 
Here, $\vec s$ is the spin of electrons and $\lambda$ is a phenomenological constant. In Eq. (\ref{ap2}), $\vec r_i=\vec 0$ and $\vec R$ for the first and second localized moments, respectively. 
Localized moment located at $\vec R$ interacts with the perturbed wavefunctions of the electrons, found in Eq. (\ref{ap1}), so one can write the interaction total energy as
$E =\sum\limits_{\vec k\nu}^{occ} \langle\widetilde{\vec k\nu}\mid\hat V_2\mid\widetilde{\vec k\nu}\rangle+ \langle\vec k\nu\mid\hat V_1\mid\vec k\nu\rangle$, where
$occ$ stands for occupied and shows the fact that only occupied states will be taken into account. Writing this energy in the form of $E=E_0+E(\vec R)$, we have
%
%\begin{equation}
%E =\sum\limits_{\vec k\nu}^{occ} \langle\widetilde{\vec k\nu}\mid\hat V_2\mid\widetilde{\vec k\nu}\rangle+ \langle\vec k\nu\mid\hat V_1\mid\vec k\nu\rangle=E_0+E(\vec R)
%\label{ap3}
%\end{equation}
%
%where
%
\begin{align}
 & E_0=\sum\limits_{\vec k\nu}^{occ} \langle\vec k\nu\mid\hat V_1+\hat V_2\mid\vec k\nu\rangle,     \nonumber \\
 & E(\vec R)=\sum\limits_{\vec k\nu}^{occ} \langle\vec k\nu\mid\hat V_2\,\hat G\,\hat V_1\mid\vec k\nu\rangle + h.c.=T+T^*,
\label{ap3}	
\end{align}
%
%
%Assuming $n$ and $N$ as the dimension of the space and number of sites, respectively, and 
%
where $E_0$ is a constant shift to the energy of the system $E(\vec R)$ is the $\vec R$-dependent part of the total energy. For the systems without any broken symmetries or the cases that 
summation over all the spin of electrons is zero, in other words the system is not spin-polarized, $E_0$ will vanish. Based on this fact, we only focus on $E(\vec R)$ 
which is the more interesting part of the total energy.
Using the completeness relations: $\sum\limits_{\sigma}\int d^n r=\sum\limits_{\vec r\sigma}\mid\vec r\sigma\rangle\langle\vec r\sigma\mid=1$ and 
$\sum\limits_{\vec k\sigma}\mid\vec k\sigma\rangle\langle\vec k\sigma\mid$ $=$ $\sum\limits_{\vec k\nu}\mid\vec k\nu\rangle\langle\vec k\nu\mid=1$,
and defining $\psi_{\vec k\nu}(\vec r,\sigma)=\langle\vec r\sigma\mid\vec k\nu\rangle$ we find
\begin{equation}
T =\lambda^2\sum\limits_{\vec k\nu}^{occ} \sum\limits_{\sigma_1\sigma_2}
\psi_{\vec k\nu}^*(\vec R,\sigma_1)\langle\sigma_1\mid\vec S_2\cdot\vec s \,\, G(\vec R,0,\varepsilon_{\vec k\nu})\, \vec S_1\cdot\vec s\mid\sigma_2\rangle\,\psi_{\vec k\nu}(0,\sigma_2).
\label{ap4}
\end{equation}
In the above equation, $G(\vec R,0,\varepsilon_{\vec k\nu})$ is the retarded Green's function matix evaluated at $E=\varepsilon_{\vec k,\nu}$
which its elements can be found using
\begin{equation}
G_{\sigma_1\sigma_2}(\vec r_1,\vec r_2,E)=\sum\limits_{\vec k\nu}^{\infty}\frac{\psi_{\vec k\nu}(\vec r_1,\sigma_1)\psi_{\vec k\nu}^*(\vec r_2,\sigma_2)}{E+i\mu-\varepsilon_{\vec k\nu}},
\label{ap5}
\end{equation}
with $\mu\rightarrow 0^+$. 
We will now use the facts $G(\vec R,0,\varepsilon_{\vec k\nu})=\int_{-\infty}^{\infty}\,G(\vec R,0,E)\,\delta(E-\varepsilon_{\vec k\nu})$
and $\int_{-\infty}^{\infty}   \  dE  \times   \sum_{\vec k\nu}   \rightarrow    \int_{-\infty}^{E_F}    \ dE  \times \sum_{\vec k\nu}^{occ}$
%\begin{equation}
%G(\vec R,0,\varepsilon_{\vec k\nu})=\int_{-\infty}^{\infty}\,G(\vec R,0,E)\,\delta(E-\varepsilon_{\vec k\nu}) dE
%\label{eq7}
%\end{equation}
to write $T$ in the form of 
\begin{equation}
T =\lambda^2\int_{-\infty}^{E_F}dE \sum\limits_{\sigma_1\sigma_2}
\langle\sigma_1\mid\vec S_2\cdot\vec s \, G(\vec R,0,E)\, \vec S_1\cdot\vec s\mid\sigma_2\rangle\,
\sum\limits_{\vec k\nu}^{\infty}\psi_{\vec k\nu}^*(\vec R,\sigma_1)\psi_{\vec k\nu}(0,\sigma_2)\delta(E-\varepsilon_{\vec k\nu}).
\label{ap6}
\end{equation}
Here, there is an important point: to get the final result we have changed the order of integration over $\vec k$ and $E$ in Eq. (\ref{ap6}). 
As it has been said in the paper, if Fubini's condition\cite{Fubini1,Fubini2} is not satisfied, changing the order of integration in a double-integral can change the result of
integration. It is important to emphasize that when one uses the last result of this Appendix, Eq. (\ref{master}), to find the magnetic interaction, 
it should be carefully evaluated, and in the case that the result does not make physical sense, the method of integration over $\vec k$, which is using $E(\vec R)=2 Re[T]$ with $T$ defined in the Eq. (\ref{ap4}), 
should be used.

The $\delta$-function in Eq. (\ref{ap6}) can be expressed in a form that it is useful to make Green's function. 
The equality $\lim_{\mu\to 0}(x\pm i\mu)^{-1}=P(x^{-1})\mp i\pi\delta(x)$ leads to the fact 
$\delta(x)=(i/2\pi)\,\lim_{\mu\to 0}((x+ i\mu)^{-1}-(x- i\mu)^{-1})$. 
Expanding $\delta(E-\varepsilon_{\vec k\nu})$ in Eq. (\ref{ap6}) in that form, and using 
$\sum_{\sigma} \mid \sigma\rangle \langle \sigma\mid=1$ and $E(\vec R)=2 Re[T]$ with $T$ defined in the Eq. (\ref{ap4})
result the $\vec R$-dependent part of the energy as
\begin{align}
E(\vec R)=&\,\frac{-\lambda^2}{\pi}\,\,Im \int_{-\infty}^{E_F}Tr\big[G(0,\vec R, E)\, \vec S_2\cdot\vec s\,\, G(\vec R, 0, E)\, \vec S_1\cdot\vec s\, \big]dE  \nonumber \\ 
          &+\frac{\lambda^2}{\pi}\,\,Im \int_{-\infty}^{E_F}Tr\big[G^{A}(0,\vec R, E)\, \vec S_2\cdot\vec s\,\, G(\vec R, 0, E)\, \vec S_1\cdot\vec s\, \big]dE
\label{ap7}
\end{align}
where $G^{A}(0,\vec R, E)$ is Advanced Green's function matrix which its elements can be found using 
\begin{equation}
G^{A}_{\sigma_1\sigma_2}(\vec r_1,\vec r_2,E)=\sum\limits_{\vec k\nu}^{\infty}\frac{\psi_{\vec k\nu}(\vec r_1,\sigma_1)\psi_{\vec k\nu}^*(\vec r_2,\sigma_2)}{E-i\mu-\varepsilon_{\vec k\nu}},
\label{ap8}
\end{equation}
with $\mu\rightarrow 0^+$.
It is convenient to show that second term in Eq. (\ref{ap7}) turns out to be zero. To show this, one can easily expand the Green's fucntion
matrices in terms of Pauli matrices, then use the fact that $G^{A}(0,\vec R, E)=G^{\dagger}(\vec R,0,E)$.
The last fact leads to the desired result of this Appendix,
\begin{equation}
E(\vec R)=\frac{-\lambda^2}{\pi}\,\,Im \int_{-\infty}^{E_F}
Tr\big[G(0,\vec R, E)\, \vec S_2\cdot\vec s\,\, G(\vec R, 0, E)\, \vec S_1\cdot\vec s\, \big]dE.
\label{finalmaster}
\end{equation}

\end{document}